\documentclass[twocolumn,showpacs,prc,aps,floatfix,reprint]{revtex4-1}
\usepackage[T1]{fontenc}
\usepackage[utf8]{inputenc}
\usepackage{graphicx}%
\usepackage{dcolumn}
\usepackage{bm}
\usepackage{hyperref}
\usepackage{amssymb}
\usepackage{braket}
\usepackage{graphicx}
\usepackage{dcolumn}
\usepackage{amsmath}
\usepackage{amsfonts}
\usepackage{color}
\usepackage{bm}
\usepackage{siunitx}
\DeclareUnicodeCharacter{0229}{\c{e}}
\begin{document}

\preprint{APS/123-QED}

\title{Dispersive-optical-model analysis of the asymmetry dependence of neutron skins}

\author{R. A. Ramon$^{1}$, M. C. Atkinson$^{2,3}$, and W. H. Dickhoff$^1$}

\affiliation{${}^1$Department of Physics,
Washington University, St. Louis, Missouri 63130}
\affiliation{${}^2$Department of Physics,
Reed College, Portland, OR 97202}
\affiliation{${}^3$Department of Physics and Astronomy,
Haverford College, Haverford, PA 19041}

\date{\today}

\begin{abstract}
New dispersive optical model analyses of ${}^{54}$Fe and ${}^{90}$Zr together with updated results for ${}^{40}$Ca, ${}^{48}$Ca, and the earlier result for ${}^{208}$Pb shed light on the behavior of neutron skins in nuclei. Starting with a negative skin for ${}^{40}$Ca, a trend increasing somewhat stronger than linear emerges when the neutron skin of these nuclei is considered as a function of asymmetry, $(N-Z)/A$, and linked to the Green's function Monte Carlo results for asymmetric He nuclei. 
This general trend is consistent with the expectation that nuclei near the neutron drip line are expected to have very large neutron skins. 
The present analysis therefore motivates the question of which nuclei provide the most relevant link to neutron star physics. 
\end{abstract}

\maketitle

\section{Introduction\label{sec:intro}}

The mantle of excess neutron matter in a neutron-rich finite nucleus is known as the neutron skin. Quantitatively, the difference between the point neutron and proton root-mean-squared (RMS) radii is defined as the skin, $R_\mathrm{skin} = R_n - R_p$, of a finite nucleus. The precise determination of $R_\mathrm{skin}$ is crucial to understand nuclear properties, including masses, radii, and the location of the neutron drip line, and to test nuclear models~\cite{Typel01,furnstahl2002occupation,Steiner05,RocaMaza11}.

Because the properties of finite nuclei are governed by the same underlying effective interactions that dictate infinite nuclear matter, $R_{\text{skin}}$ provides an important link to the nuclear equation of state (EOS).
However, the connection between finite nuclei and infinite nuclear matter is more complex than a straightforward extrapolation due to shell effects and surface properties in nuclei~\cite{Atkinson:2020B}.
Nevertheless, $R_\mathrm{skin}$ has been interpreted as a measure of the density dependence of the symmetry energy around saturation~\cite{Typel01, furnstahl2002occupation, Steiner05, RocaMaza11}. In particular, $R_\mathrm{skin}$ is linked to the slope of the symmetry energy $L$, since a thicker skin favors a large $L$ and a thinner skin a smaller one. 
Its importance thus extends to heavy-ion reactions~\cite{Li08} as well as to astrophysics for understanding supernovae and neutron stars~\cite{Horowitz01, Steiner10} due to the aforementioned symmetry energy role in the equation of state (EOS) for nuclear matter.

Based on these considerations, the neutron skin has been the subject of many studies, both experimental and theoretical, to determine its thickness~\cite{Tsang12, mammei2024neutron}. Experimentally, nuclear charge density distributions can be measured precisely by well-studied electromagnetic interactions~\cite{Angeli:2013, garcia2016unexpectedly}. 
It is considerably more difficult to determine the radius of the neutron density distribution, so the majority of past neutron skin measurements employed hadronic probes which have significant model dependencies~\cite{piekarewicz2006insensitivity}.
Electroweak probes~\cite{PREX12, adhikari2021accurate,adhikari2022precision} have no such limitations~\cite{PREX12,donnelly1989isospin} but are particularly difficult to perform with suitable precision due to how weak the electron-neutron interaction is. The first parity-violating experiment performed by the PREX collaboration yielded a thick neutron skin of $^{208}$Pb with a rather large uncertainty~\cite{PREX12}. A later experiment, PREX-2, resulted in a skin of $0.283 \pm 0.071$ fm~\cite{adhikari2021accurate}, more precise than the earlier one. The CREX experiment resulted in a much smaller skin with higher precision in $^{48}$Ca of $0.121 \pm 0.026$ (exp) $\pm 0.024$ (model) fm~\cite{adhikari2022precision}.

Until recently, no theoretical approach could simultaneously describe the neutron skin of $^{48}$Ca and $^{208}$Pb. 
Expectations based on mean-field models suggested a strong correlation between these two skins, although the large error bars for PREX-2 may not provide a stringent constraint on the isovector part of energy density functionals~\cite{reinhard2022combined}. Beyond the mean field, there are also \textit{ab initio} calculations for both nuclei. In Ref.~\cite{hagen2016neutron}, a neutron skin for ${}^{48}$Ca was predicted, which is consistent with CREX, while the results of Refs.~\cite{hu2022ab, hu2024author} exhibit mild tension with PREX-2.
All these approaches focus on calculating ground-state properties without addressing the implications for the scattering domain~\cite{Dussan:2014} and the approaching continuum at the drip line.
The dispersive optical model (DOM) provides an alternative approach, accurately describing both skins of $^{48}$Ca and $^{208}$Pb as well as other ground-state properties while also providing an excellent description of elastic nucleon scattering data~\cite{calleya2025investigating}.

The DOM describes scattering observables in addition to bound nucleon and ground-state properties by employing a dispersion relation that couples the energy domains above and below the Fermi energy. Employing Green’s function theory~\cite{Exposed!}, the DOM establishes the nucleon self-energy as an empirical, nonlocal optical potential constrained by both bound-state and scattering measurements~\cite{Mahzoon:2014, atkinson2020dispersive}.
Its efficacy has been confirmed by its success in predicting $(e,e'p)$ cross sections without fitting to these experimental data~\cite{Atkinson:2018,Atkinson:2019}.

In the present paper, we report on a DOM study of neutron skins including new results for ${}^{54}$Fe and ${}^{90}$Zr, updated results for ${}^{40}$Ca and ${}^{48}$Ca, and earlier results for ${}^{208}$Pb.
The goal is to address the general dependence of the neutron skin on nucleon number which requires the inclusion of the role of the continuum as it provides an obstacle to a straightforward extrapolation to infinite matter.
Difficulty with extrapolating to infinite matter was already documented for the energy per particle employing the DOM for a variety of nuclei~\cite{Atkinson:2020B}.
We note that at this time only the DOM is available to incorporate the role of the continuum as it takes its presence into account by providing an accurate description of proton and neutron elastic scattering data~\cite{Dickhoff:2017,Dickhoff:2019}.
In addition, it provides an accurate description of the charge density when sufficient elastic electron scattering data are available.

With the difficulty of obtaining precise neutron skin information, the community has investigated general trends across the nuclear chart.
A linear trend of neutron skin thickness as a function of isospin asymmetry was deduced from antiprotonic x-ray data~\cite{trzcinska2001neutron, jastrzebski2004neutron}.
Ab initio coupled-cluster theory and auxiliary field diffusion Monte Carlo methods also exhibit a linear trend of the neutron skin~\cite{novario2023trends}. These results span an isospin asymmetry from -0.30 to 0.30.
A different systematic study critiques the overall linear trend from earlier works~\cite{zhang2021systematic}. Instead of an overall single trend, Ref.~\cite{zhang2021systematic} obtains separate trends for different isotopic chains, but the results for $^{208}$Pb and $^{48}$Ca do not coincide with the neutron skins of the PREX and CREX experiments as they are only described simultaneously by the DOM~\cite {calleya2025investigating}. 

In Sec.~\ref{sec:DOM} we review the properties of the DOM relevant for the study of neutron skins.
Results for charge and weak charge densities are discussed in Sec.~\ref{sec:results} together with our findings for the neutron skin behavior implied by these results. 
We also draw attention to the importance of Green's function Monte Carlo results for ${}^6$He and ${}^8$He as these nuclei provide the only reliable results for the neutron skin at large nucleon asymmetry.
Some conclusions are drawn in Sec.~\ref{sec:conclusions}.

\section{Dispersive optical model\label{sec:DOM}}
The nucleon self-energy, considered as a complex one-body potential, can be parametrized as an optical potential with mostly standard functional forms as proposed by Mahaux and Sartor~\cite{Mahaux91}. 
 
 The present implementation of the DOM employs a subtracted dispersion relation linking the energy domain of structure to the one of elastic scattering~\cite{Exposed!}, 
 \begin{align}
     \mathrm{Re}\ \Sigma^*(\bm{r},\bm{r}';E) &= \mathrm{Re}\
     \Sigma^*_{HF}(\bm{r},\bm{r}';\varepsilon_F) \label{eq:dispersion} \\ -
     \mathcal{P}\int_{\varepsilon_F}^{\infty} \!\! \frac{dE'}{\pi}&\mathrm{Im}\
     \Sigma^*(\bm{r},\bm{r}';E')\left[\frac{1}{E-E'}-\frac{1}{\varepsilon_F-E'}\right] \nonumber
     \\ + \mathcal{P} \! \int_{-\infty}^{\varepsilon_F} \!\!
     \frac{dE'}{\pi}&\mathrm{Im}\
     \Sigma^*(\bm{r},\bm{r}';E')\left[\frac{1}{E-E'}-\frac{1}{\varepsilon_F-E'}\right],
     \nonumber      
  \end{align}
  where $\bm{r},\bm{r}'$ imply also discrete quantum numbers and
  $\varepsilon_F$ is the average Fermi energy which separates the particle and hole domains, 
 \begin{align*}
 \varepsilon_F = \frac{1}{2}\left(E_0^{A+1}-E_0^{A-1}\right),
     \end{align*}
 and $E_0^{A\pm1}$ is the ground state energy of the $A\pm1$ nucleus. 
 The energy-independent correlated Hartree-Fock (HF) contribution~\cite{Exposed!} is thus removed.
The subtracted form has the further advantage that the emphasis is placed on energies closer to the Fermi energy for which more experimental data are available.
The real part of the self-energy at the Fermi energy is still referred to as the HF term, $\Sigma^*_{HF}$~\cite{Mahaux91,Dickhoff:2017,Dickhoff:2019},  but is sufficiently attractive to bind the empirical orbits.
Initially, standard local functional forms for these terms were employed~\cite{Charity06,Charity:2007,Mueller:2011} but the inclusion of the ground-state charge density requires a fully nonlocal implementation to succeed~\cite{Dickhoff:2010,Mahzoon:2014} further supported by the theoretical work of Refs.~\cite{Waldecker:2011,Dussan:2011}.

We employ a nonlocal version of the DOM self-energy 
 where $\Sigma^*_{HF}(\bm{r},\bm{r'})$ and
  $\mathrm{Im}\ \Sigma^*(\bm{r},\bm{r'};E)$ are parametrized    and Eq.~\eqref{eq:dispersion} generates the energy dependence of the
  real part~\cite{atkinson2020dispersive,calleya2025investigating,Mahzoon:2014}. 
 To employ the DOM self-energy for predictions, the parameters are fit through a weighted $\chi^2$ minimization of available elastic differential cross section data ($\frac{d\sigma}{d\Omega}$), analyzing power data ($A_\theta$), reaction cross sections ($\sigma_r$), total cross sections ($\sigma_t$), charge density ($\rho_{\text{ch}}$), energy levels ($\varepsilon_{\ell j}$), particle number, separation energies, the root-mean-square charge radius ($R_\mathrm{ch}$), and the energy of the ground state~\cite{Atkinson:2020B,atkinson2020dispersive}. The scattering calculations are performed using the framework of $R$-matrix theory~\cite{Baye:2010,Baye_review} and the bound-state calculations utilize the Green's function formalism~\cite{Exposed!}.
 In the following we only present those quantities that are directly relevant for the results of the present paper.

For the results below the Fermi energy, 
we employ the Dyson equation in a partial wave basis (with orbital angular momentum $\ell$ and total angular momentum $j$) for either protons or neutrons to obtain the Green's function, $G_{\ell j}(r,r';E)$, from the DOM self-energy, 
  \begin{align}
    G_{\ell j}(r,r';E) &= G_{\ell}^{(0)}(r,r';E)
    \label{eq:dyson} \\ +
    \int \!\! dr_1 dr_2 r_1^2 r_2^2 &G_{\ell}^{(0)}(r,r_1;E)\Sigma_{\ell
    j}^*(r_1,r_2;E)G_{\ell j}(r_2,r';E) ,
     \nonumber
 \end{align}
 where $G^{(0)}_{\ell}(r,r';E)$ corresponds to the free propagator (the Green's function when $\Sigma_{\ell j}^*(r_1,r_2;E)=0$)
 ~\cite{Exposed!}. The particle number, binding energy, and charge density are all obtained from the hole spectral function which corresponds to the imaginary part of the Green's function, 
 \begin{equation}
    S_{\ell j}(r,r';E) = \frac{1}{\pi}\mathrm{Im}\ G_{\ell j}(r,r';E) .
    \label{eq:spec}
 \end{equation}
The single-particle density distribution can be calculated from the hole spectral function according to
 \begin{equation}
    \label{eq:charge}
    \rho(r) = \frac{1}{4\pi} \sum_{\ell j} (2j+1) \int_{-\infty}^{\varepsilon_F}dE\ S_{\ell j}(r,r;E) .
 \end{equation}
The RMS radii of the proton and neutron distributions of Eq.~(\ref{eq:charge}) are used to calculate $R_\mathrm{skin}$ as well as the nuclear charge radius,
\begin{equation}
    \label{eq:rch}
    R_\mathrm{ch}^2 = R_p^2 + \braket{r_p^2} + \frac{N}{Z}\braket{r_n^2} + \braket{r_\mathrm{DF}^2} + \braket{r_{SO}^2}, 
 \end{equation}
 where $\braket{r_{SO}^2}$ is the spin-orbit contribution calculated according to Ref.~\cite{atkinson2020dispersive}, $\braket{r_p^2} = 0.709$ fm$^2$ is the charge radius squared of the proton~\cite{pohl2010size}, $\braket{r_n^2} = -0.106$ fm$^2$ is the charge radius squared of the neutron~\cite{filin2020extraction}, and $\braket{r_\mathrm{DF}^2}$ is the so-called Darwin-Foldy term which is a relativistic correction.
 To obtain the charge density, $\rho_\mathrm{ch}(r)$, we fold the single-particle densities with neutron and proton charge distributions
in addition to calculating their spin-orbit contributions as detailed in Refs.~\cite{atkinson2020dispersive,Horowitz:2012,calleya2025investigating}.

\section{Results\label{sec:results}}

\begin{figure}[b]
    \includegraphics[width=\columnwidth]{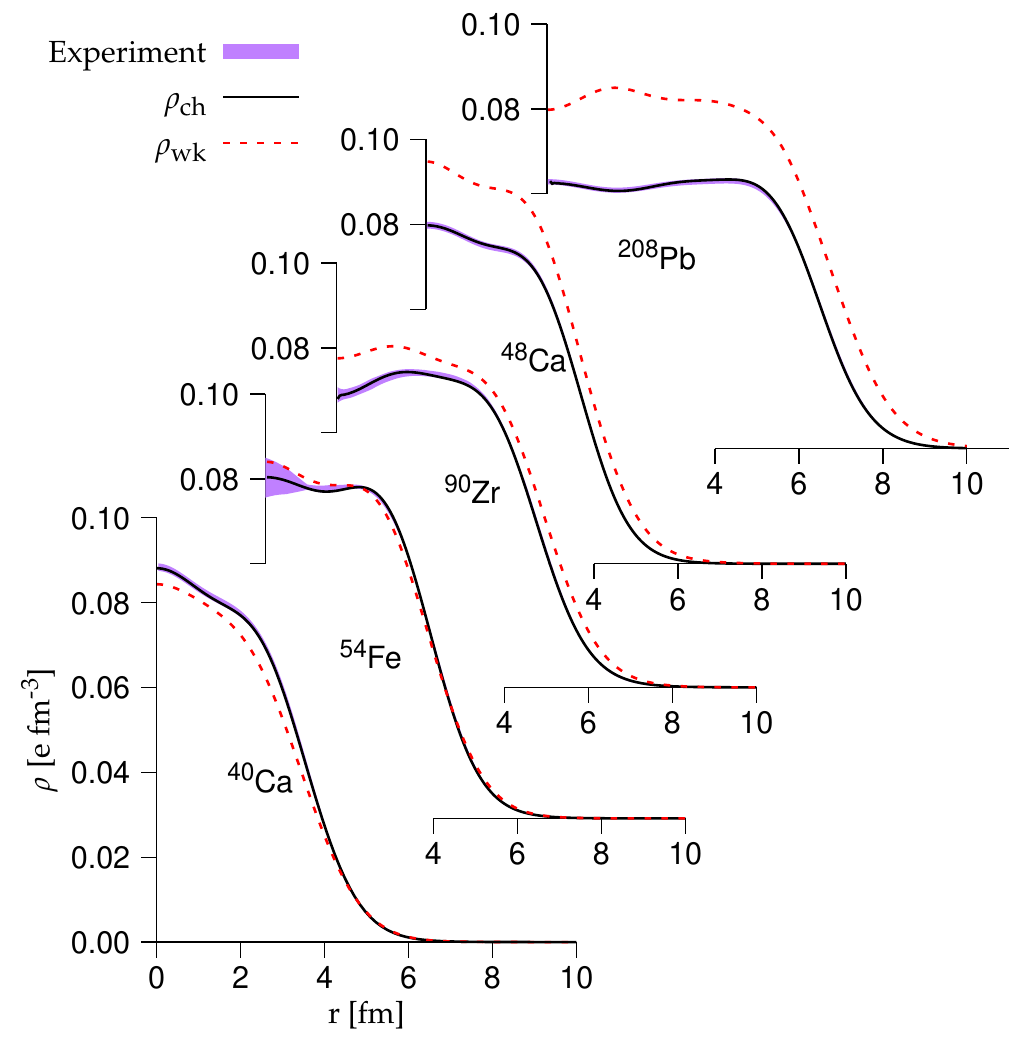}
    \caption{Ground-state charge and weak charge density distributions calculated in the DOM framework. The shaded purple bands represent the experimental results and associated errors for the charge density. Solid lines represent the DOM calculated charge density, $\rho_{\textrm{ch}}$, and red dashed lines identify the weak charge distribution, $\rho_{\textrm{wk}}$. The label $\mathrm{e}$ on the vertical axis therefore refers either to the charge or the weak charge. These distributions are ordered according to increasing isospin asymmetry.}
   \label{fig:Combined_chd}
\end{figure}

New DOM results have been generated for ${}^{90}$Zr that extend the reach of the DOM to semi-magic nuclei by including the effect of proton-proton pairing near the Fermi energy~\cite{Ramon:2026B}.
This nucleus has been studied by the Nikhef group, providing accurate high-resolution $(e,e'p)$ cross sections~\cite{den1988single}.
The analysis presented in Ref.~\cite{Ramon:2026B} yields an excellent description of these data including the effect of the proton pairing. 
The $1p\frac{1}{2}$ spectroscopic factor \textit{e.g.} corresponds to 0.4 and its partial contribution facilitates an accurate description of the experimental charge density shown in Fig.~\ref{fig:Combined_chd}. 
These results support the notion that the DOM is capable of describing all experimental data of semi-magic nuclei that are related to the single-particle propagator of the system~\cite{Exposed!}.

Updated results for ${}^{40}$Ca and ${}^{48}$Ca are available that provide an improved treatment of high-momentum components but otherwise do not change the results for ground-state densities~\cite{Ramon:2026A}.
While the small neutron skin of ${}^{48}$Ca was correctly described in Ref.~\cite{calleya2025investigating}, the required reduction of the neutron radius had the effect of generating more neutron high-momentum components than for protons.
This result disagrees with the experimental observation that the minority species should dominate the high-momentum regime over that of the majority species~\cite{Duer:2018} as predicted by \textit{ab initio} calculations of asymmetric matter~\cite{Frick:2005,Rios:2009,Rios:2014}.
The DOM high-momentum spectral functions obtained in Ref.~\cite{Mahzoon:2014} also exhibit a defect in that the peak of the distribution doesn't move to higher missing energy with higher missing momentum.
Both issues were resolved by extending the DOM to include volume absorption below the Fermi energy with the radius decreasing with increasing missing energy~\cite{Ramon:2026A} in agreement with \textit{ab initio} work reported in Refs.~\cite{Muether:1995,Dussan:2011}.
This extension abandons the factorization of the energy dependence and geometry of the potential as originally proposed by Mahaux and Sartor~\cite{Mahaux91}. 
The resulting charge densities are displayed in Fig.~\ref{fig:Combined_chd} and are equivalent to earlier results in Ref.~\cite{Atkinson:2018} and \cite{Atkinson:2019,calleya2025investigating}, respectively.

The numerical extension of the DOM has been employed to generate a DOM description of ${}^{54}$Fe~\cite{Ramon:2026C} to study the recent surprising results of the CaFe experiment~\cite{Nguyen2026}.
The corresponding charge density for this open-shell proton system also exhibits excellent agreement with the experimental result as shown in Fig.~\ref{fig:Combined_chd}.

As mentioned earlier, we include in our analysis of ground-state densities these new results for these four nuclei and the earlier result for ${}^{208}$Pb that successfully predicted the neutron skin of this nucleus~\cite{atkinson2020dispersive}.
The resulting electric and weak charge densities are displayed in Fig.~\ref{fig:Combined_chd} organized according to increasing asymmetry $(N-Z)/A$.
Such ordering identifies the increasing extent of the weak charge distribution in comparison to the charge distribution.
Experimental charge densities were obtained from Ref.~\cite{deVries:1987} and are shown as shaded areas, the solid lines show the DOM description of charge distributions, and the dashed lines represent the neutron dominated weak charge distribution for each nucleus.
All values for the charge radii are accurate to within 0.5\% for all nuclei. 
Results for the neutron skin are -0.06 fm for ${}^{40}$Ca, -0.04 fm for ${}^{54}$Fe, 0.08 fm for ${}^{90}$Zr, 0.12 fm for ${}^{48}$Ca, and 0.25 fm for ${}^{208}$Pb, respectively, with an error of about 20\% for ${}^{208}$Pb 
 (see Refs.~\cite{Mahzoon:2017,atkinson2020dispersive} for the uncertainty quantification).
The skin results are in agreement with the recent PREX~\cite{adhikari2021accurate} and CREX~\cite{adhikari2022precision} results and are based on significant data sets of other experimental properties for both protons and neutrons that are accurately described by the DOM analysis.

\begin{figure*}[t]
    \includegraphics[width=0.49\textwidth]{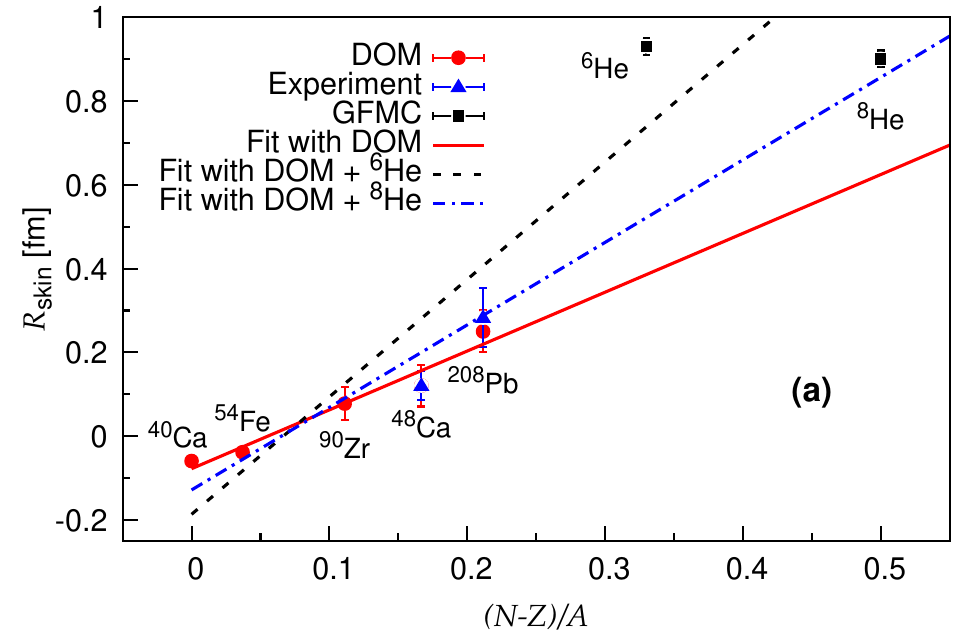}
    \includegraphics[width=0.49\textwidth]{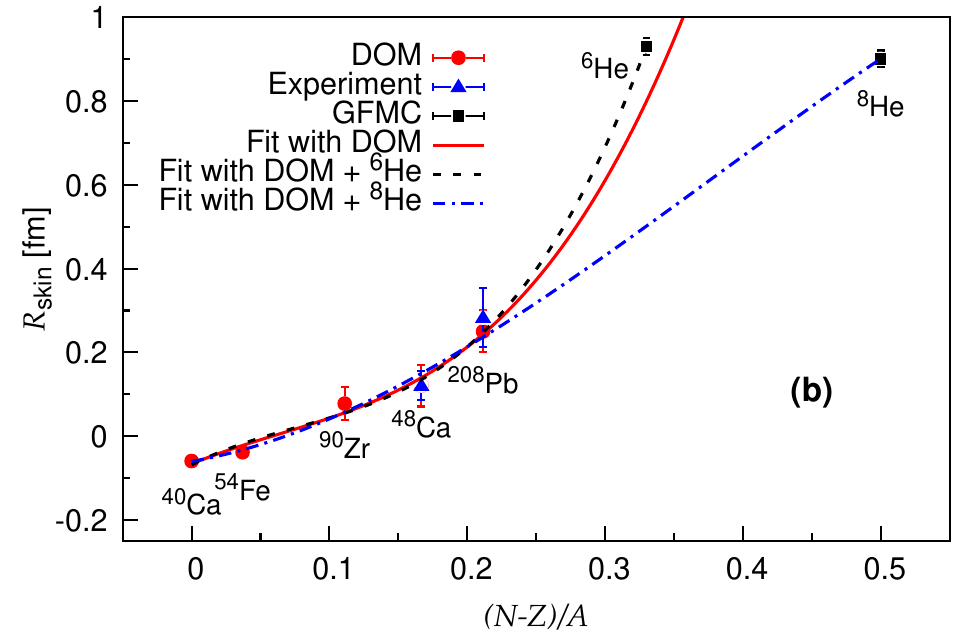}
    \caption{Neutron skin as a function of asymmetry, $(N-Z)/A$. The circles indicate DOM, and the squares represent GFMC results. The triangles are from experiments, CREX and PREX-2 for $^{48}$Ca and $^{208}$Pb, respectively. Black dashed, and blue dash-dot lines represent the fitted curves for DOM \& $^{6}$He and DOM \& $^{8}$He, respectively. The red solid line shows the curve fitted only to the DOM results. Panel (a) shows linear fits, and (b) shows $3^{rd}$ order polynomial fits.}
   \label{fig:skin_trend}
\end{figure*}

These nuclei have asymmetry within the region studied in previous analyses~\cite{zhang2021systematic,ozawa2001nuclear}, thus a linear fit (using nonlinear least-squared regression) between $R_\textrm{skin}$ and nucleon asymmetry describes all of the DOM results, as demonstrated by the solid red line in Fig.~\ref{fig:skin_trend}. 
The DOM skin results (red circles) are in agreement with the recent PREX~\cite{adhikari2021accurate} and CREX~\cite{adhikari2022precision} results displayed in Fig.~\ref{fig:skin_trend} by blue triangles.

We now draw attention to theoretical information that provides the only other reliable information about neutron skins.
The Green's function Monte Carlo (GFMC) method is capable of exactly solving for the nuclear ground states (and some excited states) of light nuclei for a given realistic Hamiltonian containing two- and three-body interactions. 
Results for the neutron skins of $^{6}$He and $^{8}$He are quoted in Ref.~\cite{Carlson:2015} which extends the nucleon asymmetry range up to $(N-Z)/A=0.50$. 
These GFMC results represent exact calculations for the combined two- and three-body Hamiltonian that yield agreement with the experimental charge radii and given the correct description of nuclei up to $A$ = 12, we assume these neutron skin results to be reliable and shown as black squares in Fig.~\ref{fig:skin_trend}. 
These results are confirmed by the GFMC calculation~\cite{Maria:2026} reported in Ref.~\cite{Piarulli:2018} with a different Hamiltonian~\cite{Maria:2016}.

When considering these neutron skin results for asymmetric He isotopes, none of the linear fits in Fig.~\ref{fig:skin_trend}(a) sufficiently cover the data, regardless of whether we include only DOM points in the fit (the red solid line), the DOM points and $^{6}$He (black dashed line), or the DOM points and $^8$He (blue dash-dotted line). Therefore, the trend appears to be somewhat stronger than linear when the entire asymmetry range is considered and these He isotopes are included. 

\begin{figure}[b]
    \includegraphics[width=\columnwidth]{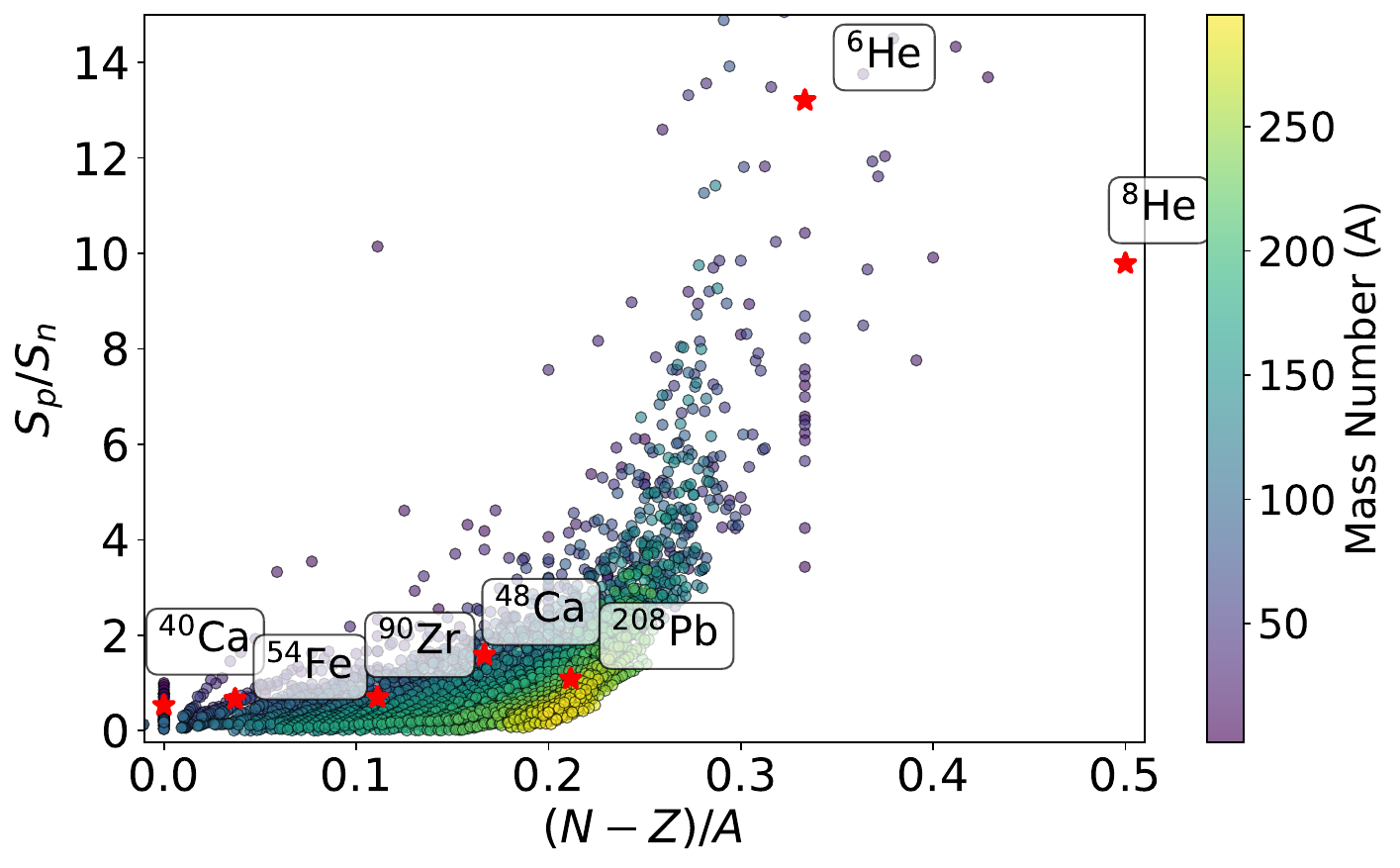}
    \caption{Change of the ratio of proton and neutron separation energy, $(S_p/S_n)$ with isospin asymmetry $(N-Z)/A$. Red stars indicate nuclei relevant for this study. Separation energies are collected from the National Nuclear Data Center~\cite{NNDC}. For better visualization, only the range $0\leq{S_p}/{S_n}\leq15$ was plotted, which covers $99.1\%$ of isotopes with non-negative ${S_p}/{S_n}$ and asymmetry.}
   \label{fig:ratio_asymmetry}
\end{figure}

It is useful to consider the analysis of Ref.\cite{zhang2021systematic} which found a resemblance between the patterns of $S_p/S_n$ \textit{vs}. $(N-Z)/A$  and $R_\mathrm{skin}$ \textit{vs}. $(N-Z)/A$. Furthermore, a correlation between $R_\mathrm{skin}$ and the difference between proton and neutron separation energies, $(S_p-S_n)$, is reported in~\cite{ozawa2001nuclear} with a range from $-15$ to $30$ MeV.  
The ratio, $(S_p/S_n)$, measures the relative binding of protons and neutrons. Naturally, neutron-rich nuclei have smaller neutron separation energies compared to protons, resulting in a larger $(S_p/S_n)$ ratio. These weakly bound neutrons result in a neutron density distribution that is expected to be more extended than the proton density distribution. 
This behavior is illustrated in Fig.~\ref{fig:ratio_asymmetry}, where the change of $(S_p/S_n)$ as a function of $(N-Z)/A$ for all available data of neutron-rich nuclei is plotted. 
We have highlighted  ${}^{40}$Ca, ${}^{48}$Ca, ${}^{54}$Fe, ${}^{90}$Zr, and ${}^{208}$Pb in Fig.~\ref{fig:ratio_asymmetry} which have been analyzed with the DOM in Refs.~\cite{Mahzoon:2017,calleya2025investigating,atkinson2020dispersive,Ramon:2026B,Ramon:2026C}. 
In addition, we identify ${}^{6}$He and ${}^{8}$He with asymmetries 0.33 and 0.50, respectively, for which GFMC results for the neutron skin are available~\cite{Carlson:2015}.
A complete picture of the neutron skin properties of nuclei must include weakly bound systems such as these. 
The nonlinear trend displayed in Fig.~\ref{fig:ratio_asymmetry} together with the plausible connection to $R_\mathrm{skin}$ suggests that the complete relationship between neutron skin and nucleon asymmetry may well be nonlinear. 

Since a linear fit in Fig.~\ref{fig:skin_trend}(a)is insufficient, and the ratio $(S_{p}/S_{n})$ in Fig.~\ref{fig:ratio_asymmetry} also suggests a non-linear trend with asymmetry, we explored higher-order fits. A quadratic fit was similarly insufficient, but a cubic fit provides a possible description of the data, as seen in Fig.~\ref{fig:skin_trend}(b). The black dashed line in Fig.~\ref{fig:skin_trend}(b), which is constrained by the DOM results and $^{6}$He, is in good agreement with all data points, excluding $^8$He. It is remarkable that the fit constrained by only DOM points (the solid red line in Fig.~\ref{fig:skin_trend}(b)), with asymmetry between 0 and 0.21, predicts a $^6$He neutron skin (with asymmetry 0.33) very close to the GFMC result. This suggests that the trend between $R_\mathrm{skin}$ and $(N-Z)/A$ is stronger than linear. 
A similar fit including DOM results and ${}^8$He but excluding ${}^6$He is also successful.
Considering the wide range of data in Fig.~\ref{fig:ratio_asymmetry} we conjecture that many of these nuclei will exhibit a neutron skin within the boundaries of the dashed and dash-dotted curve in Fig.~\ref{fig:skin_trend}(b) with some extending even further.

The He isotopes illustrate the sensitivity of the neutron skin to shell effects with the four extra neutrons in ${}^8$He appearing to compress the ${}^4$He more than the two extra neutrons in ${}^6$He as the charge radius is also smaller in ${}^8$He~\cite{mueller:2007}. 
We note that ${}^8$He is also an outlier in Fig.~\ref{fig:ratio_asymmetry} and one would expect weakly bound neutron systems to exhibit a variety of large neutron skins depending on nuclear structure effects.
This notion is consistent with the multiple trends observed in  Ref.~\cite{ozawa2001nuclear}; in Fig.~\ref{fig:skin_trend}(b) we see a trend that contains $^6$He and another trend that contains $^8$He. This again emphasizes the importance of finite-nucleus effects in determining neutron skins.    

\section{Conclusions}
\label{sec:conclusions}

While our DOM results do not cover rare isotopes, Fig.~\ref{fig:skin_trend}(b) suggests that a wide spread of increasing neutron skins with asymmetry is to be expected, thereby rendering the increase in the neutron skin from ${}^{48}$Ca to ${}^{208}$Pb much less surprising and consistent with expectations for nuclei containing very weakly bound neutrons near the neutron drip line.
The presence of the continuum in finite nuclei provides an important impediment to relating information about neutron skins in nuclei to properties of neutron stars.
Consequently, determining which specific finite nuclei provide the most robust constraint on astrophysical EOS models remains an open question. 
Resolving this issue will require theoretical frameworks beyond standard mean-field approaches that can systematically incorporate continuum couplings and simultaneously account for high-precision benchmarks like PREX and CREX.
\vspace{1em}
\begin{acknowledgments}
This work was supported by the U.S. National Science Foundation under grants PHY-2207756 and PHY-2512895.
\end{acknowledgments}

\bibliographystyle{apsrev4-1}
\bibliography{Skin_2026}

\end{document}